\documentclass[twocolumn,showpacs,preprintnumbers,amsmath,amssymb,prb]{revtex4}
\usepackage{graphicx}
\usepackage{dcolumn}
\usepackage{bm}
\begin{document}
\title{Quantum Knitting Computer}

\author{Toshiyuki Fujii}
\author{Shigemasa Matsuo}
\author{Noriyuki Hatakenaka}%

\affiliation{Graduate School of Integrated Arts and Sciences, Hiroshima University, 
Higashi-Hiroshima, 739-8521, Japan.}

\date{\today}

\begin{abstract}
We propose a fluxon-controlled quantum computer incorporated with three-qubit quantum error correction using special gate operations, i.e., joint-phase and SWAP gate operations, inherent in capacitively coupled superconducting flux qubits. The proposed quantum computer acts exactly like a knitting machine at home. 
\end{abstract}
\maketitle

The physical implementation of quantum computation requires a series of accurately controllable quantum logic gates 
consisting of quantum-mechanical coherent manipulations on a single qubit and an arbitrary pair of qubits. 
The gate controls are in general complicated and lead to computation errors of decoherence as a result of coupling with the ambient environment during quantum state evolution, especially on a large scale. Here we propose a simple gate-control scheme designed to achieve quantum computation 
incorporated with quantum error correction by using a single fluxon moving along a Josephson transmission line (JTL). 

The system we are considering is composed of a Josephson transmission line and a zigzag chain of bipartite superconducting flux qubits \cite{fluxqubit} 
with an alternating arrangement, i.e., one qubit functions as a data qubit, while its nearest neighbor functions as a switch between data qubits (see Fig. 1 (a)) 
\cite{Matsuo, Matsuo2}. The $1$st basic block of the chain is shown in Fig. 1(b). In general, the $i$-th basic block is composed of two data qubits labeled d$i$, and d$(i+1)$ and one switch qubit labeled s$i$. For simplicity, the energy-level separations of all the data qubits are assumed to be equal, 
and very different from those of switch qubits. Thus, the data qubits are initially decoupled from each other in this system. 

Let us consider the situation where a fluxon approaches the $i$-th switch qubit. 
The coupling of two data qubits is realized via through the $i$-th switch qubit, which is controlled by a fluxon motion, i.e., the qubit-qubit interaction is turned on when 
the fluxon induces an energy-level shift equal to the energy-level separation of data qubits so as to resonate energetically among three qubits. 
During the resonance, the three qubit state vector $|\Psi (t) \rangle_i$ of the $i$-th basic block evolves through the relation 
$
|\Psi (t) \rangle_i = U_i(t) |\Psi(0)\rangle_i
$,
with $U_{i}(t)$ being a time-translational operator for the $i$-th basic block. A significant example for quantum computation is the transfer of information 
in the basic block of the chain \cite{Matsuo, Matsuo2}. In particular, the state vector starting from the switch qubit state $|{0}\rangle^s_i$ and the arbitrary two-data qubit state $| {\psi} \rangle^d_{i,i+1}$ is decoupled again with $\pi$ pulse application, i.e., 
$U_i(t_\pi ) \{ |\psi \rangle^d_{i, i+1} \otimes|{0}\rangle^s_i \} 
= \{U_i |{\psi (0)}\rangle^d_{i, i+1 }\}\otimes |{0}\rangle^s_i$ 
where $t_\pi=\hbar\pi/g\sqrt{2}$ with $g$ being the coupling constant. The effective time-translational operator for the data qubit in the $i$-th basic block system 
$U^0_i$ is then expressed in a matrix form as 
\begin{equation}
U^0_i =
\begin{pmatrix}
1&& 0 &&0&& 0 \\
0&&0 && -1 &&0 \\
0 && -1 &&0 &&0 \\
0&&0 &&0&& -1 \\
\end{pmatrix}\label{is}
\end{equation} 
where the basis is ordered as $ |00\rangle^d_{i,i+1}$, $ |01\rangle^d_{i,i+1}$, $ |10\rangle^d_{i,i+1}$, $ |11\rangle^d_{i,i+1}$. 
\begin{figure}
\begin{center}
\includegraphics[width=6cm,clip]{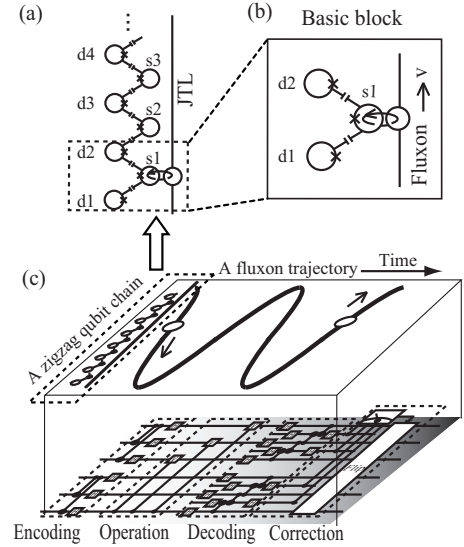}
\caption{\label{EB} 
Quantum knitting computer: 
(a) proposed quantum circuit, 
(b) basic block of  circuit, 
(c) schematic diagram of  quantum information knitting. }
\end{center}
\end{figure}
This is regarded as a gate composed of a joint-phase gate, which reverses its sign only in the $|{00}\rangle^d_{i, i+1}$ state, and a SWAP gate, sometimes called a JP+SWAP(JPS) gate \cite{Yung}, 
\begin{equation}
U^0_i =
- 
\begin{pmatrix}
-1&& 0 &&0&& 0 \\
0&&1 && 0 &&0 \\
0 && 0 &&1 &&0 \\
0&&0 &&0&& 1 \\
\end{pmatrix} \begin{pmatrix}
1&& 0 &&0&& 0 \\
0&&0 && 1 &&0 \\
0 && 1 &&0 &&0 \\
0&&0 &&0&& 1 \\
\end{pmatrix}. \label{is}
\end{equation}

The universal gate for quantum computation that we use here is a controlled-NOT (CNOT) plus SWAP gate, called a CNS gate \cite{schuch2003ntq, benjamin2004qca}. As shown in Fig. 2 (a), this can be formed by performing single-qubit operations such as flip (X) and Hadamard (H) operations on the JPS gate before and after {\it one-way} fluxon propagation. 

By using CNS gates, an arbitrary two-qubit controlled operation that induces an arbitrary unitary operation parametrized by four real values $\alpha$, $\beta$, $\delta$ and $\theta$, 
\begin{equation}
V=e^{i \delta} R_z (\alpha ) R_y (\theta ) R_z (\beta ), \label{u}
\end{equation}
with a rotation operator 
$
R_i (\theta ) = e^{-i \theta \sigma_i /2 } =I \cos{\theta /2} - i\sigma_i \sin{\theta /2} \label{ri}
$
to the target bit, say qubit 3, if and only if the control bit, say qubit 1, is in the state $|1\rangle_1^d, $ can be accomplished as shown in Fig. 2  (b). 
Here $\sigma_i $, $(i=x,y,z)$ is a Pauli operator. 
The operators $A$, $B$, and $C$ are defined as
\begin{equation}
\begin{cases}
A &= R_z(\alpha ) R_y \left( \frac{\theta }{2} \right) \\
B &= R_y \left(-\frac{\theta }{2}\right) R_z \left(-\frac{\alpha +\beta }{2} \right) \\
C &= R_z \left(\frac{\beta -\alpha }{2} \right). 
\end{cases} \label{abc}
\end{equation}
This includes two-time CNS gates, therefore, {\it round-trip} fluxon propagation with certain single-qubit operations leads to an arbitrary type of two-qubit controlled operation in this basic block of qubits.

\begin{figure}
\begin{center}
\includegraphics[width=6cm,clip]{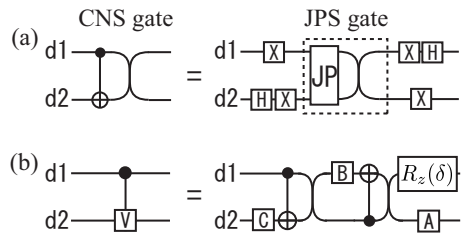}
\caption{\label{fp} 
(a) Circuit diagram of a CNS gate created by {\it one-way} fluxon propagation 
together with single-qubit operations. 
H and X are a Hadamard gate and a bit flip gate, respectively. JP stands for a joint-phase gate.
(b) Circuit diagram for a controlled operation with an arbitrary unitary transformation $V$ caused 
by {\it round-trip} fluxon propagation. }
\end{center}
\end{figure}

To excute quantum computation in a zigzag chain consisting of a number of the basic blocks of qubits as shown in Fig. 1 (a), 
a controlled operation between an arbitrary pair of qubits, which are in general apart from each other, should be established even in a system with only a nearest-neighbor interaction. Fortunately, quantum-state transfer is possible due to the JPS operation inherent in our chain without the need for any other complicated gate controls \cite{Matsuo}. In other words, the qubit information can move along the zigzag chain, just like a movable qubit acting as a quantum data bus accompanying a fluxon. 
Then the passage of a {\it one-way} fluxon through the chain together with single-qubit operations generates CNS operations between the data qubit d1 and all other data qubits as shown in Fig. 3 (a). Note that single-qubit operations are performed in all the qubits simultaneously before and after the fluxon passes through the entire chain. The {\it round-trip} fluxon propagation together with operations \eqref{abc} in the chain 
provides a controlled operation with an arbitrary unitary transformation between the control qubit and the target qubit that are apart from each other 
as shown in Fig. 3 (b). This operation can also be applied to multi-qubits simultaneously, resulting in short excution times. 

\begin{figure}
\begin{center}
\includegraphics[width=6cm,clip]{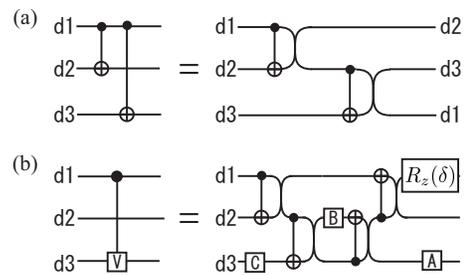}
\caption{\label{fCU} 
(a ) Three data qubit operation realized by {\it one-way } fluxon propagation.
(b) Controlled operation between data qubit d1 and distant data qubit d3 realized by {\it round-trip } fluxon propagation.}
\end{center}
\end{figure}

The control qubit can be assigned by a single-gate control. By introducing a gate bias to the switch qubits connected to the desired data qubit to achieve off-resonance when the fluxon passes through the chain, the swap operation accompanying the fluxon can be interrupted at the desired data qubit. This qubit now acts as a starting qubit. 
In this way, the fluxon with single-qubit controls acts as a cam box or carriage passing across a bed of needles causing the needle movements required to produce successive stitches in a knitting machine. 

The controlled NOT operations between a particular qubit and other qubits 
are also useful as regards quantum error correction as shown in Fig. 3 (a). 
Quantum error correction schemes have been developed 
using the redundant multi-qubit encoding of quantum data 
combined with error-detection and recovery steps. 
No error-correction scheme has yet been implemented in the solid-state qubits 
required for complicated switch controls. 

The dominant noise in a superconducting qubit is attributed to dephasing, namely a loss of phase coherence between the computational bases in the system. 	
This type of error (noise) causes randomly reversing the sign of the encoded qubit's state (a phase-flip error), i.e., $| \phi \rangle^d = c_0 | 0 \rangle^d + c_1 |1\rangle^d \rightarrow \sigma_z |\phi \rangle^d = c_0 | 0 \rangle^d - c_1 |1\rangle^d $ with probability $p$. 
It is expected that there will be no more than two data-bit errors in the system simultaneously if $p$ is sufficiently small. 
Such an error can be corrected by using a three-qubit code \cite{preskill, shor, steane, nielsen2000qca} that uses the three data qubits to represent a logical qubit. 
This error correction code has been successfully implemented in other systems \cite{cory, chiaverini} proving the possibility of prolonging the lifetime of a logical qubit.
Moreover, Josephson transmission lines acting as good high-pass filters contribute to 
the attenuation of low frequency noise below the plasma frequency, resulting in the suppression of the phase noise caused by switching. 
In addition, the encoding and error detection processes can realized by {\it a few } fluxon propagations. We use the neighboring $i$-th and $(i+1)$-th basic blocks as the $i$-th logical qubit, then we encode the logical qubit state $|\phi \rangle_{i}^L$ on the three data qubits using a new set of bases $|0 \rangle_{i}^L \equiv |+\rangle^d_{i}\otimes | +\rangle^d_{i+1} \otimes |+ \rangle^d_{i+2} $ and $|1 \rangle^L_i \equiv |-\rangle^d_{i} \otimes | -\rangle^d_{i+1} \otimes |- \rangle^d_{i+2}$, 
where $|\pm \rangle^d_i = |0 \rangle^d_i \pm |1 \rangle^d_i $ \cite{nielsen2000qca}. This encoding process is accomplished by {\it one-way} fluxon propagation as shown in Fig. 4. 
A phase flip error is represented by the bit flip in the bases $|\pm \rangle^d$, 
i.e, $\sigma_z | \pm \rangle^d = | \mp \rangle^d $. 

\begin{figure}
\begin{center}
\includegraphics[width=8cm,clip]{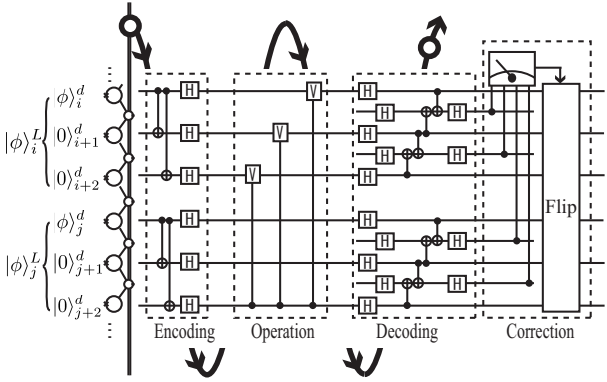}
\caption{\label{ec} Quantum knitting computer with error correction.}
\end{center}
\end{figure}

The flipped qubit in the $i$-th logical qubit block is detected without destroying coherence as follows. 
We first transform the each data qubit $|\pm \rangle^d$ 
to $|0 \rangle^d$ and $|1 \rangle^d$ 
by operating the Hadamard gate for all data qubits, 
and then all switch qubits are also transformed to 
$|0 \rangle^s_{i} \rightarrow |{0}\rangle^s_{i} +|{1}\rangle^s_{i}$. 
After this single qubit operation, 
a fluxon propagates along the chain to braid the quantum information.
Then the {\it one-way } fluxon propagation changes all the qubits as, 
\begin{equation}
\begin{split}
U_i(t_\pi ) \left\{ |{\psi}\rangle^d_{i, i+1} \otimes \left( |{0}\rangle^s_i +|{1}\rangle^s_i \right) \right\} \hspace{-3.5cm}& \\
&= \left\{ U_0^i |{\psi}\rangle^d_{i, i+1} \right\} \otimes |{0} \rangle^s_i + \left\{ U_1^i |{\psi} \rangle^d_{i, i+1} \right\} \otimes |{1} \rangle^s_i \\
 &=\left\{ U_+^i|{\psi}\rangle^d_{i, i+1} \right\} \otimes |{+} \rangle^s_i + \left\{ U_{-}^i |{\psi}\rangle^d_{i, i+1} \right\} \otimes |{-} \rangle^s_i, \label{EDetect}
\end{split}
\end{equation}
where $U^{\pm}_i=U^0_i \pm U^1_i$ with $U^1_i$ being the unitary operator 
for two neighboring data qubits under the switch qubit state $|{1}\rangle^s_i$, 
i.e., 
$ U_i(t_\pi ) \{| \psi \rangle^d_{i,i+1} \otimes | 1 \rangle^s_i \} 
= \{ U_1^i | \psi \rangle^d_{i,i+1} \} \otimes | 1 \rangle^s_i $, 
expressed as 
\begin{equation}
 U^1_i=-
\begin{pmatrix}
1&& 0 &&0&& 0 \\
0&&0 && 1 &&0 \\
0 && 1 &&0 &&0 \\
0&&0 &&0&& -1 
\end{pmatrix}.\label{bis}
\end{equation} 
Therefore, $U^{\pm}_i$ are given as 
\begin{eqnarray}
U^+_i &=&
-|{01}\rangle \langle{10} |_{i,i+1}^d - |{10} \rangle \langle{01}|_{i,i+1}^d 
\label{u+} \\
U^-_i &=&|{11} \rangle \langle {11}|_{i,i+1}^d +|{00} \rangle \langle {00}|_{i,i+1}^d. 
\label{u-}
\end{eqnarray} 
The $U^+_i $ and $U^-_i$ operators examine whether the states of the neighboring data qubits are the same. If the error has occurred at one of the two data qubits, 
the neighboring qubits have different values, leading to $|+ \rangle^s_i$ in the switch qubit from Eq. \eqref{u+}. 
On the other hand, the lack of any difference in the data qubits is confirmed by detecting $|- \rangle^s_i$ in the switch qubit from Eq. \eqref{u-}. 
To identify an error location, we need to the results with those for the neighboring $(i+1)$-th basic block qubits as shown in Table \ref{Ecor}. 
Finally the detected error is corrected by flipping the incorrect qubit. 

\begin{table}
\vspace{0.5cm}
\begin{tabular}{|c|c|}\hline
 $(i, i+1)$-th switch qubit &Flipped data qubit \\
\hline
 $(+, +)$ & $ {i+2}$\\
$(+, -)$ & ${i+1}$\\
$(-, +)$ & ${i}$\\
$(-, -)$ & \text{None}\\ \hline
\end{tabular}
\caption{\label{Ecor} The error bit corresponding to the measured switch qubit values. }
\end{table}

In summary, we have investigated quantum computation in a zigzag qubit chain running alongside a Josephson transmission line. 
A special gate control such as JP+SWAP gate is inherent in our proposed system. This allows us to simplify quantum computation. 
The gate controls can only be performed by fluxon round-trip propagation together with single qubit controls, 
just like a knitting machine at home. In addition, this gate also enables us to perform quantum error correction naturally. 
Thus our system provides quantum computation incorporated with quantum error correction that is difficult to achieve with existing solid-state qubits. 
Moreover, a fluxon train consisting of multi-fluxons may make it possible to increase the execution rate. 
\begin{acknowledgments}
We thank M. Fukunaga for fruitful discussions. 
This work was supported in part by KAKENHI (Nos. 18540352 and 195836) from MEXT of Japan.
\end{acknowledgments}

\bibliographystyle{unsrt}

\end{document}